\documentclass{ws-mpla}

\begin{document}

\markboth{G\'eraldine Servant}
{STATUS REPORT ON UNIVERSAL EXTRA DIMENSIONS
AFTER LHC8}

%%%%%%%%%%%%%%%%%%%%% Publisher's Area please ignore %%%%%%%%%%%%%%
\catchline{}{}{}{}{}
%%%%%%%%%%%%%%%%%%%%%%%%%%%%%%%%%%%%%%%%%%%%%%%%%%%%%%%%%%%%%%%%%%%

\title{STATUS REPORT ON UNIVERSAL EXTRA DIMENSIONS AFTER LHC8
}

\author{\footnotesize G\'eraldine SERVANT}

\address{Instituci\'o Catalana de Recerca i Estudis Avan\c{c}ats (ICREA) \&
Institut de F\'isica d'Altes Energ\`ies (IFAE), 
Universitat Auton\`oma de Barcelona,
08193 Bellaterra, Barcelona, Spain\\
gservant@ifae.es}

\maketitle

%\pub{Received (Day Month Year)}{Revised (Day Month Year)}

\begin{abstract}
Although they do not address the hierarchy problem, models with Universal Extra Dimensions 
have attracted a lot of attention 
as simple benchmark models characterized by small mass splittings and a dark matter WIMP played by the Lightest Kaluza-Klein particle (LKP).  
We review their status, with emphasis on the minimal implementation in 5 dimensions (MUED) in which the LKP is a massive hypercharge gauge boson.  In this case, the mass range accounting for the correct dark matter abundance (around 1.4 TeV) remains untouched by LHC8 and is out of reach of present DM direct detection experiments. However, LHC14 can probe the relevant region in the 3-lepton channel.

\end{abstract}

%\ccode{PACS Nos.: include PACS Nos.}

\section{Introduction}	

Models with Universal Extra Dimensions (UED) \cite{Appelquist:2000nn} are probably the simplest extra-dimensional models.
%and have been the subject of many phenomenological studies. 
The same gauge symmetry and particle content as  the Standard Model is embedded in 5 or 6 flat space-time dimensions. 
Their distinctive properties follow from  non-trivial assumptions about the boundary Lagrangians which respect a space-time symmetry, called Kaluza-Klein (KK) parity. 
In particular, while in most extra-dimensional models, Kaluza-Klein  states are not stable  as they are all able to decay into SM particles, in  UED, the Lightest Kaluza-Klein particle is stable as a consequence of KK parity.

Fifteen years ago, extra dimensions at the TeV scale were motivated by the hierarchy problem, similar to Supersymmetry when the superpartner masses are around a TeV. 
%Both large extra dimensions a la ADD and warped extra dimensions have the potential to address the %hierarchy problem of the SM.
However, unlike warped extra dimensions for instance, 
UED models do not address the hierarchy problem. 
The Higgs particle is added by hand, except in Ref.~\cite{ArkaniHamed:2000hv},  which shows that in the case that  the Standard Model gauge forces propagate in 6 or 8 dimensions, the Higgs scalar is automatically generated as quark composites, bound by the Standard Model gauge forces which become strong in the bulk.  In contrast with the usual four-dimensional dynamical electroweak symmetry breaking  models,  the binding force can be the SM gauge interactions themselves, without the need of introducing new strong interactions. 
Anyhow, these scenarios generally share the same problems as models where the Higgs is a $t\overline{t}$ condensate, the predicted Higgs mass is typically too large  and the 
low energy effective theory requires tuning (see Ref.~\cite{Cheng:2013qwa} for a related recent discussion).
Another attempt to address the little hierarchy problem in UED was provided in \cite{Burdman:2006jj} by embedding either a Little Higgs or Twin Higgs model in UED.
 Moreover, to fully address the hierarchy problem requires a mechanism that dynamically selects the compactification scale to be near the TeV scale, to connect the compactification scale and electroweak symmetry breaking (Ref.~\cite{Appelquist:2002ft} proposes to do so by adding a seventh warped extra dimension).

UED models should be contrasted to 5D {\it Gauge-higgs unification} models~\cite{Panico:2006em} in which 
  the Higgs field is the internal component of a 5D gauge field, whose mass is protected by 5D gauge invariance. In that case,  the two localized lagrangians at the orbifold fixed points are different so that there is no KK parity. Consequently, EW precision constraints impose a significantly stronger bound on the KK mass scale than in UED. 
  Assuming KK parity and KK number conservation may be considered rather artificial~\cite{Agashe:2007jb}, but it leads to a distinctive phenomenology which is interesting to study in its own. For this reason, UED phenomenology has been the subject of many studies, ignoring the theoretical shortcomings of the underlying construction.

UED models have been reviewed in \cite{Hooper:2007qk}. 
Since then, there were new developments in particular regarding the relic density 
calculation\cite{Belanger:2010yx} and  LHC  bounds \cite{Belyaev:2012ai}.
 In this report,  we collect the recent updated constraints and relevant plots in view of the 14 TeV run at the LHC.
We focus on the Minimal Universal Extra Dimensions model (MUED) in which the only free parameters are the radius $R$ of the extra dimension and the cutoff scale $\Lambda$. We allude to other constructions in the last section.

\section{Basic properties of UED}

In models with Universal Extra Dimensions, 
 the Standard Model (SM) is promoted to 1+(3+n) Minkowski space-time, unlike string motivated constructions where the SM is localized on a 3-brane. 
 A new dimension potentially leads to a new conserved momentum along the extra dimension.
 Translation
invariance along an extra dimension $y$ is only broken by the orbifold projection imposed to recover a chiral SM spectrum. 
  Within exactly flat extra dimensions, the Kaluza-Klein wave functions are sines/cosines\footnote{The flatness of profiles in UED is not natural. A model of dynamical symmetry breaking in UED would typically spoil the flatness of the Higgs profile.} $\sim  \sin ny/R, \cos ny/R$ and tree-level KK masses given by $(n^2/R^2+m_0^2)^{1/2}$.
  Effective interactions are given by integrals  of wave functions along the extra dimension, leading to the  very special selection rule
 \begin{equation}
 n_1\pm n_2\pm ... \pm n_N=0 
 \label{eq:rule}
\end{equation}
for an interaction involving $N$ states with KK level $n_i$ respectively. This has a very important consequence,  that at tree-level, odd KK-number states can only be pair-produced, while a second-level  KK mode can be singly produced.
Besides, the bulk theory has an exact remnant discrete symmetry, $(-1)^n$, where $n$ is the KK number, called KK parity, which treats  different KK modes differently. 
 This symmetry insures that interaction vertices cannot involve an odd number of odd-KK states and, therefore, a vertex with two SM particles (with $n=0$) and one KK state (with $n=1$) is forbidden. As a result, the Lightest KK Particle (LKP) with $n=1$ cannot decay into SM particles and is stable. 
Note that KK parity is a reflection about the midpoint of the extra dimension combined with the orbifold projection. For it to be an exact symmetry, one has to assume that the boundary lagrangians at the two orbifold  fixed points are symmetric. 
The minimal UED  (MUED) framework assumes that boundary operators are symmetric and that the coefficients of the localized operators vanish at tree level. As a consequence of these two assumptions, KK parity is an exact symmetry and corrections to EW observables arise at 1-loop order. This results in a rather weak bound on the KK mass scale compared to other extra-dimensional constructions 
\cite{Appelquist:2000nn,Appelquist:2002wb}, namely $R^{-1}\gtrsim  680$ GeV\cite{Gogoladze:2006br,Kakuda:2013kba}. Besides, 
 the flavor structure of the bulk lagrangian, being the same as the SM one, includes the GIM mechanism.
Flavor physics constraints and other low-energy probes of new physics, such as rare $K$ and $B$ decays and muon (g-2) were summarized in Ref.~\cite{Hooper:2007qk}. 
The quoted bounds are  of the order $R^{-1}\gtrsim 250 $ GeV. 
Constraints from $b\rightarrow s \gamma$ were first derived in 
\cite{Agashe:2001xt} and from flavor changing neutral currents in \cite{Buras:2002ej,Buras:2003mk}. 
A stronger bound $R^{-1}\gtrsim 600 $ GeV from $\bar{B}\to X_s \gamma$  was subsequently derived in \cite{Haisch:2007vb}, putting flavor constraints at the same level as EW precision tests.

\begin{figure}[!b]
\centerline{\psfig{file=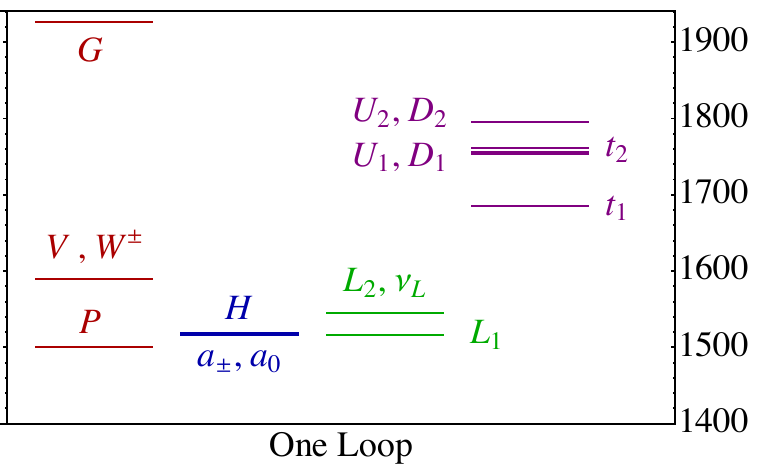,width=3.4in}}
\vspace*{8pt}
\caption{MUED 1-loop mass spectrum of level-1 Kaluza-Klein particles for $m_h=120$ GeV, $R^{-1}=1500$ GeV and $\Lambda R=20$.
From Ref.~\protect\cite{Belyaev:2012ai}.
\protect\label{figspectrum}}
\end{figure}
Generally, the
spectrum of KK masses depends also on the values of boundary terms at the cut-off scale, which are not fixed by known SM physics. In this sense, the values of the KK masses can be taken arbitrary and  the UED scenario has a multitude of parameters.
Assuming vanishing boundary terms (this is the so-called MUED hypothesis), the KK mass splittings are essentially due to radiative corrections \cite{Cheng:2002iz} and controlled by SM gauge couplings.  For a light higgs ($m_h\sim 120$ GeV), the LKP is the KK hypercharge gauge boson $B^{(1)}$ \cite{Cheng:2002iz} which behaves as a viable  alternative WIMP dark matter candidate\cite{Servant:2002aq,Cheng:2002ej} comparable to the neutralino in SUSY\cite{Cheng:2002ab}. The corresponding KK spectrum is shown in Fig.~\ref{figspectrum} in the case $R^{-1}=1.5$ TeV.

There is an additional degree of freedom which is missing in Fig.~\ref{figspectrum} and which is usually ignored in the UED literature: the radion, the scalar component of the higher dimensional graviton tensor.
The  natural scale of the radion's mass~\cite{Kolb:2003mm}  is  $m_r \sim (R^{-1})^2/m_{Pl}$. Thus, for $1/R$ at the TeV scale, the mass of the radion is at the millieV scale.
Models with flat TeV extra dimensions generally have an overclosure problem, because the radion  is effectively stable and easily dominates the energy density of the universe at late times, as extensively discussed in Ref.~\cite{Kolb:2003mm}. 
It remains to be explored whether the compactification dynamics could make the radion more massive than the naive estimate to evade the cosmological constraints. This is the assumption that is implicitly made in all the literature related to UED as the radion physics is neglected.
In the following, we will focus on dark matter and LHC constraints on UED models, ignoring the radion.

%\section{ Constraints from EW precision tests}

\section{Dark Matter}

\begin{figure}[!t]
\centerline{\psfig{file=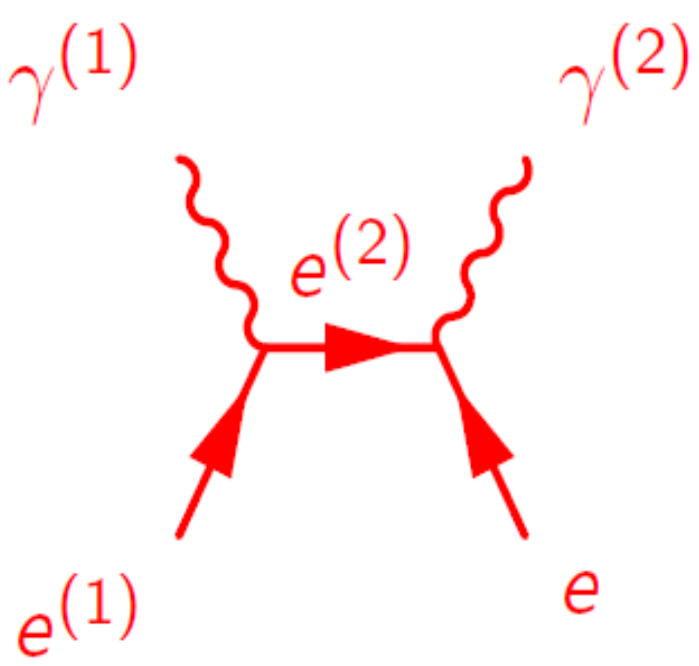,width=1.3in}}
\vspace*{8pt}
\caption{Dominant coannihilation process controlling the DM relic abundance in MUED.
%From Ref.~\protect\cite{Belanger:2010yx}. 
%Ref.~\cite{Belanger:2010yx} 
\protect\label{figdiagram}}
\end{figure}
In contrast with supersymmetry where the mass spectrum is largely spread so that at most a few additional particles participate in coannihilation processes with the LSP, in MUED, the mass spectrum of KK particles is rather degenerate and there are many coannihilation processes.

 The viability and relic density of the $B^{(1)}$ LKP were first analyzed in  \cite{Servant:2002aq} with some simplifying assumption about the KK spectrum
 (only one co-annihilation channel involving the KK right-handed electron  $e_R$ was considered) and  it was shown that including coannihilation effects increases the relic density, thus reduces the mass for the DM particle. This is to be contrasted with the supersymmetric case where coannihilation  effects tend to push the prediction for the mass of the neutralino to higher values. The reason is that the annihilation cross section of the KK photon is not helicity-suppressed and coannihilation 
 cross sections (involving KK $e_R$) are typically weaker than the ones for KK photon self annihilation and cannot compensate the increase in the number of effective degrees of freedom. 
 In Ref.  \cite{Kong:2005hn,Burnell:2005hm}, all coannihilation channels with KK fermions and KK gauge bosons were included.
The net result is that even if the new coannihilations are Boltzmann suppressed their effect is still significant because  the cross sections are mediated by weak or strong interactions while the cross sections studied so far were purely hypercharge-mediated processes.
The conclusion was that in MUED, the LKP mass should be within 500-600 GeV while in non-minimal UED models, freedom in the KK mass spectrum allows an LKP as heavy as 2 TeV\footnote{The effect of KK gravitons and the corresponding bounds on the reheating temperature were discussed in Ref.~\cite{Shah:2006gs}}.
% The effect of coannihilation with the KK Higgs was studied in 
%\cite{Matsumoto:2005uh}. 

\begin{figure}[!t]
\centerline{\psfig{file=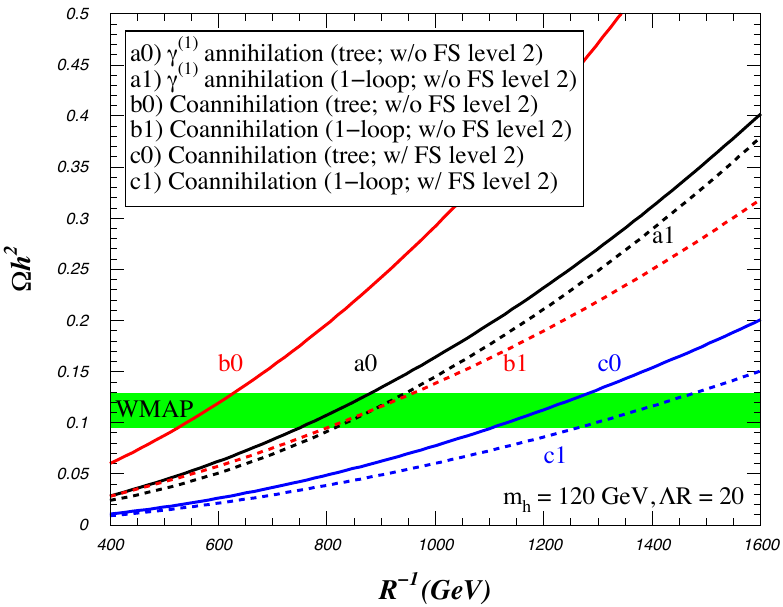,width=3.5in}}
\vspace*{8pt}
\caption{
$\Omega h^2$ predictions in MUED calculated by micrOMEGAs, taking into account EW symmetry breaking effects as well as level-2 KK particles (blue lines) as specified in framed caption. From 
Ref.~\protect\cite{Belanger:2010yx}. 
\protect\label{figrelic}}
\end{figure}

Another important subsequent result was the inclusion of level-2 KK modes which significantly increases the effective annihilation cross section \cite{Kakizaki:2005en,Kakizaki:2005uy}.  The dominant process contributing to the effective annihilation cross section (at $\sim 50$ $\%$) is $l^1 \gamma^1 \to l \gamma^2$, as shown in Fig.~\ref{figdiagram}.
At the end, the prediction for the DM mass from the relic abundance calculation in MUED is $\sim 1400$ GeV for $\Lambda R=20$, to be compared to $\sim 900$ GeV when the effect of level-2 is ignored.
The most complete calculation of the relic abundance of the KK photon was performed  in 
Ref.~\cite{Belanger:2010yx}  which showed that the prediction for the DM mass is raised to $\sim$ 1.3 TeV when taking into account the effect of level-2 KK modes. All coannihilation channels, including level-2 particles were included. The final result is shown in Fig.~\ref{figrelic} (see also Fig.~\ref{figdirect} for low cutoff scale $\Lambda=5/R$ predictions).
Interestingly, as we discuss next, the relevant mass range has not yet been reached by the LHC but will be probed in the next 14 TeV run.
\begin{figure}[!t]
\centerline{\psfig{file=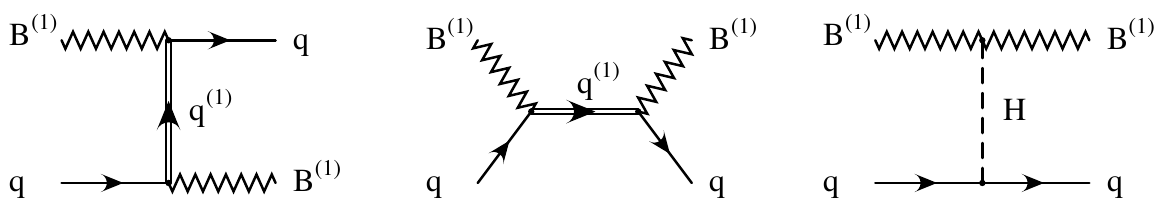,width=5.in}}
\vspace*{8pt}
\caption{Leading Feynman graphs for effective $B^{(1)}$-quark scattering through
the exchange of a KK quark (both $q^{(1)}_L$ and $q^{(1)}_R$) and through the exchange 
of a zero-mode Higgs boson. 
%The diagrams for a $Z^{(1)}$ LKP are similar.
\protect\label{fig:scattering_diagrams}}
%\vspace*{-20pt}
\end{figure}
\begin{figure}[!h]
%\centerline{\psfig{file=Figures/direct.pdf,width=3.0in}}
\centerline{\psfig{file=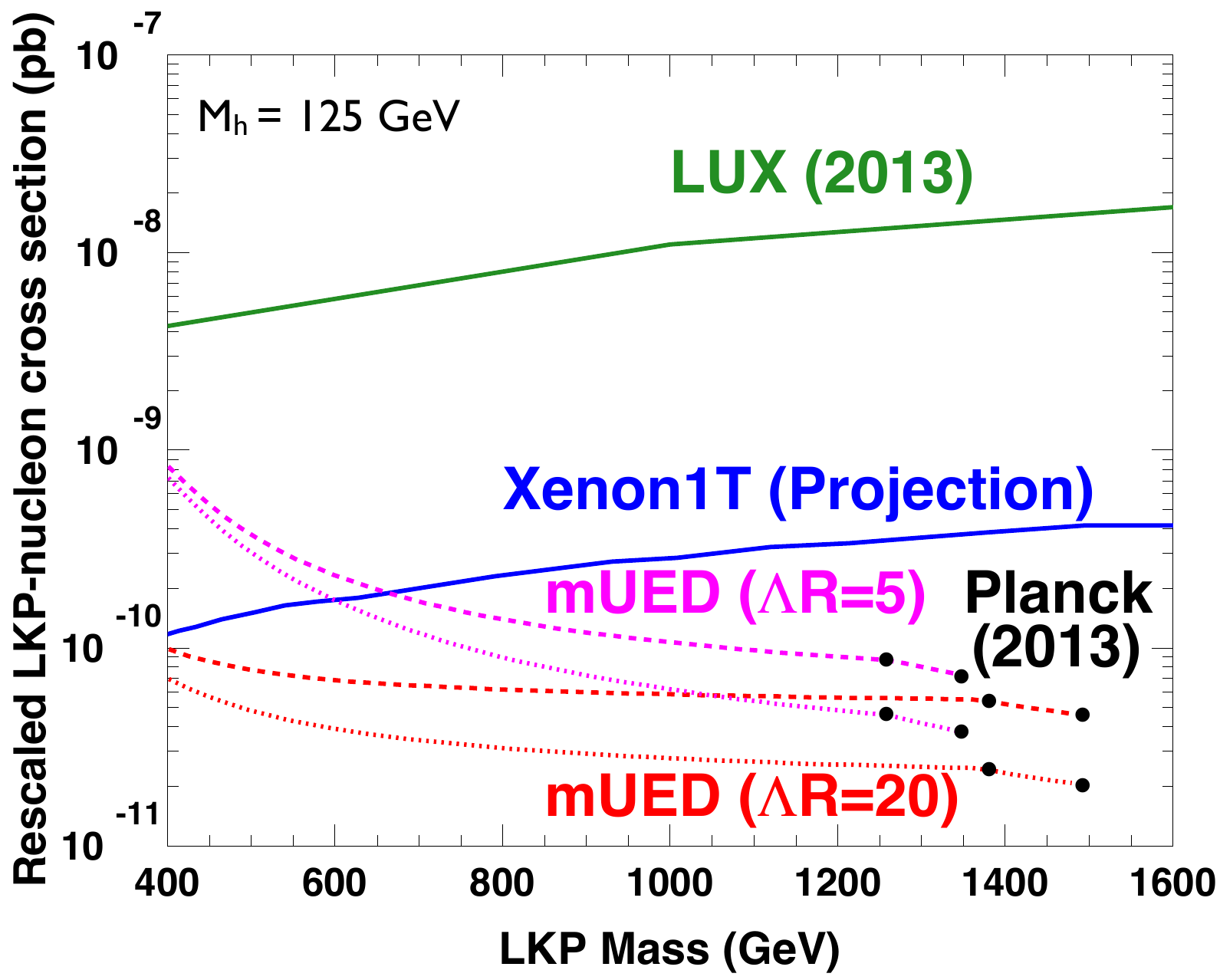,width=3.7in}}
\vspace*{8pt}
\caption{Rescaled LKP-nucleon spin-independent elastic scattering cross section $\xi \times \sigma_{SI}$ (in red) where
$ \xi = \Omega_{\mbox{\tiny LKP}} h^2/0.1103$ for $\Omega_{\mbox{\tiny LKP}} h^2 < 0.1103$,
$ \xi = 1$ for $0.1103 < \Omega_{\mbox{\tiny LKP}}h^2 < 0.1289$,
$ \xi = 0$ for $\Omega_{\mbox{\tiny LKP}}h^2 > 0.1289$, compared with the most stringent LUX 2013 bound  (green) and the Xenon 1 T projection (blue).
The region between the bullets is consistent with the $\Omega_{\mbox{\tiny DM}} h^2$ range at $3\sigma$ from the latest Planck 2013 measurement. Plotting the rescaled LKP-nucleon cross section accounts for
the fact that experimental bounds are weaker if the LKP makes only a fraction of the total dark matter energy density of the universe.
Figure provided by M. Kakizaki, using {\tt micrOMEGAs}\_3.5.5, updated from Ref.~\protect\cite{Belanger:2010yx} (which includes level-2 KK modes and loop level couplings), for two choices of cutoff scale. The dotted and dashed 
red lines correspond to two different choices of quark coefficients in the nucleon,
 see Ref.~\protect\cite{Belanger:2010yx}. 
\protect\label{figdirect}}
\end{figure}
%\subsection{Indirect detection prospects}
%

Concerning direct DM searches, they are presently insensitive to TeVish LKP.
As shown in Fig.~\ref{fig:scattering_diagrams}, elastic scattering of the $B^{(1)}$ LKP and target nuclei    arises from KK quark exchange and higgs exchange \cite{Servant:2002hb,Cheng:2002ej,Arrenberg:2008wy}, the latter being the dominant contribution for the typical mass difference of the MUED scenario, $(m_{q^1}-m_{\gamma^1})/m_{\gamma^1}\sim 17\%$. Direct detection does not appear the most promising way to probe  $B^{(1)}$ LKP dark matter as the sensitivity of near future experiments does not allow to probe realistic mass splittings leading to DM-nucleon spin-independent elastic scattering cross sections below $10^{-10}$ pb. The situation  is summarized in Fig.~\ref{figdirect}. For  a very small mass splitting between the LKP and the KK quark of the order of a couple of percents (as could occur outside of the MUED scenario by tuning appropriately bulk and/or boundary mass terms), the spin-independent cross section can rise up to a few times $10^{-9}$ pb\footnote{This is also discussed  in Ref.~\cite{Arrenberg:2013paa}, where,  on the other hand, the effects of KK modes other than $q^1$ (and especially the level-2 KK modes) are not taken into account in the relic density calculation.}. The computation of  Fig.~\ref{figdirect} includes approximate 1-loop contributions to scalar-type effective coupling with gluons from the first-level KK particles.  One-loop
contributions from KK particles higher than level-1 are neglected as
they almost decouple from the low energy effective theory. For LKP masses above 1 TeV, this calculation reproduces with a good approximation the cross section obtained 
when including more complete one-loop contributions as presented in Ref.~\cite{Hisano:2010yh,Nagata:2012xi}.

Indirect detection through gamma-rays
\cite{Cheng:2002ej,Bertone:2002ms,Bergstrom:2004cy,Baltz:2004ie,Bergstrom:2004nr,Bertone:2010fn}, neutrinos and synchrotron flux
\cite{Bertone:2002ms}, positrons \cite{Cheng:2002ej,Hooper:2004xn}, antiprotons \cite{Barrau:2005au} or through antideuterons \cite{Baer:2005tw} has also been considered.
The neutrino spectrum from LKP annihilation in the Sun was investigated in
\cite{Hooper:2004xn}.  An interesting feature of KK dark matter is, in constrast with the neutralino, that annihilation into fermions is not helicity suppressed and there can be a direct annihilation into $e^+ e^-$ leading to a  positron signal from LKP annihilation into the galactic halo  \cite{Cheng:2002ej}.  
There is presently no constraint on LKP dark matter from indirect searches for a KK DM mass of $\sim 1.4 $ TeV.

\section{LHC constraints}

%\subsection{Constraints from Higgs searches}
%\subsection{Others}
%
\begin{figure}[!t]
\centerline{\psfig{file=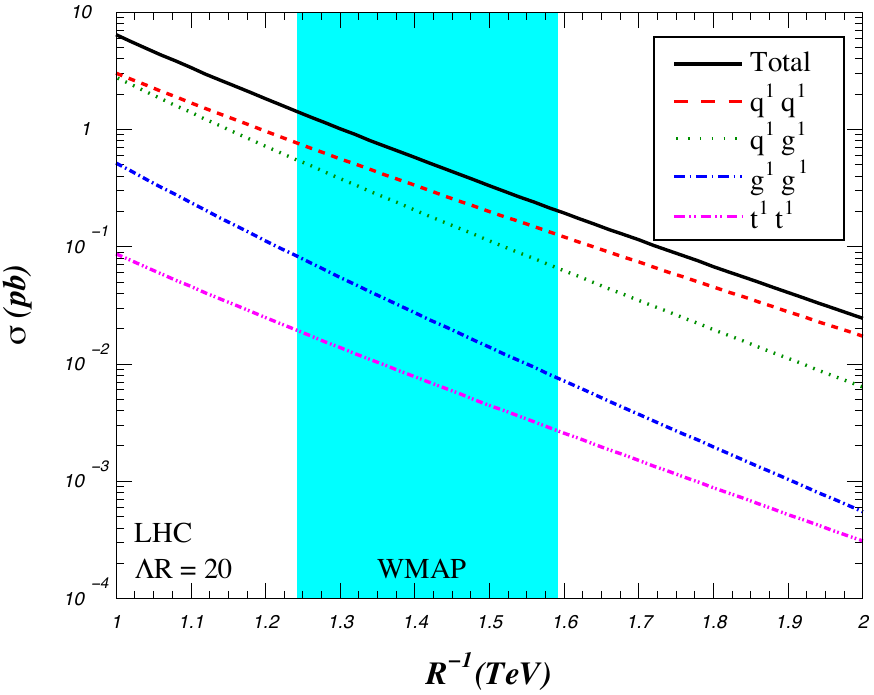,width=3.3in}}
\vspace*{8pt}
\caption{Cross section of strongly produced KK quarks and gluons at LHC14. The shaded region corresponds to the WMAP $3\sigma$ preferred region. From 
Ref.~\protect\cite{Belanger:2010yx}. 
\protect\label{figcrosssections}}
\end{figure}
\begin{figure}[!t]
\centerline{\psfig{file=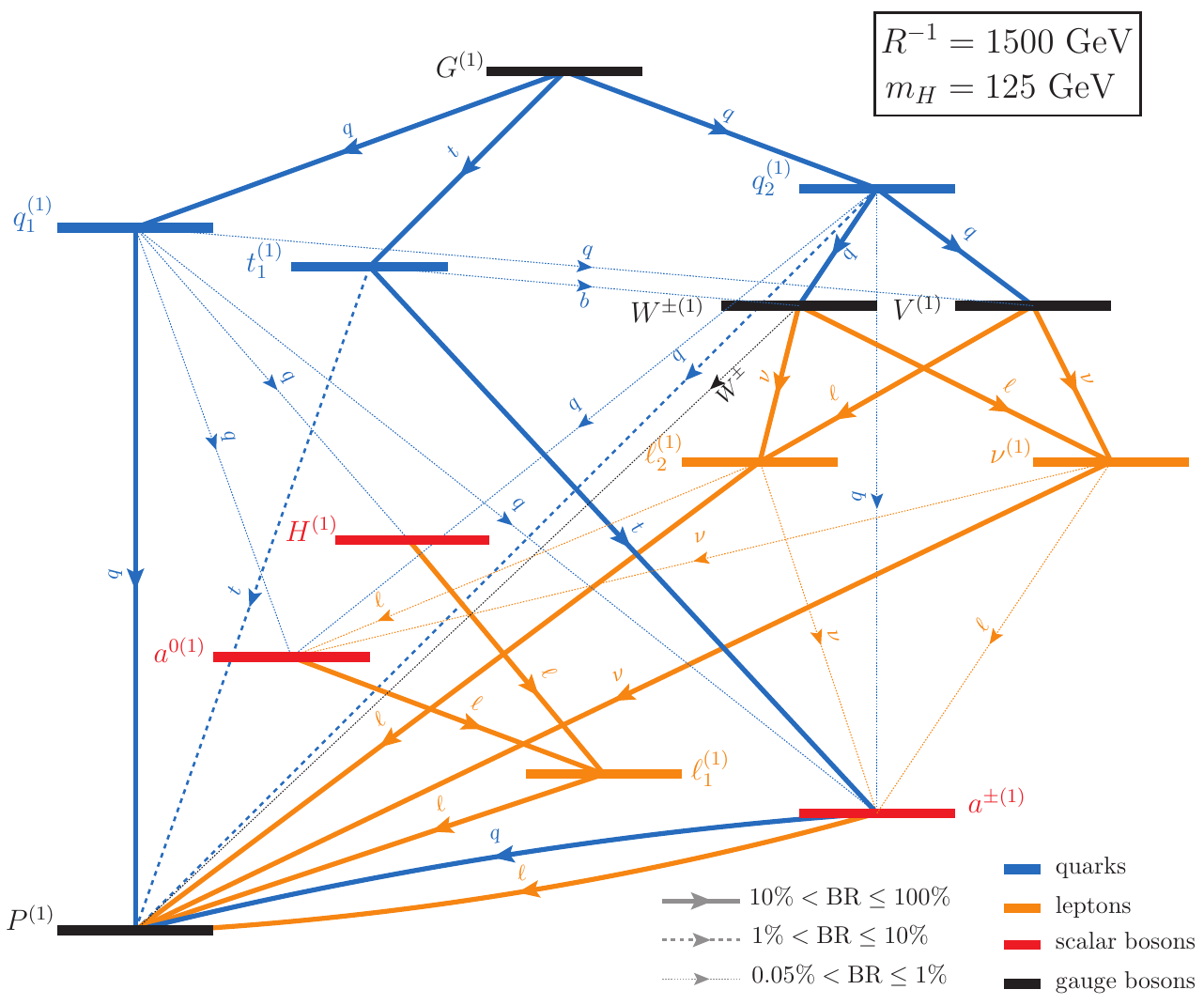,width=4.4in}}
\vspace*{8pt}
\caption{Main decay channels of the level-1 KK gluon and its decay products for $R^{-1}=$ 1.5 TeV, from  Ref.~\protect\cite{Belyaev:2012ai}.
\protect\label{figdecaychain}}
\end{figure}
The main LHC signal comes from strongly produced KK gluons and KK quarks, whose production cross sections at 14 TeV LHC are shown in Fig.~\ref{figcrosssections}. The mass spectrum of MUED defines a very specific dominant decay pattern to leptons, in contrast to SUSY theories which have instead a high quark multiplicity.  The main decay chains are shown in Fig.~\ref{figdecaychain}. 

\begin{figure}[!t]
\centerline{\psfig{file=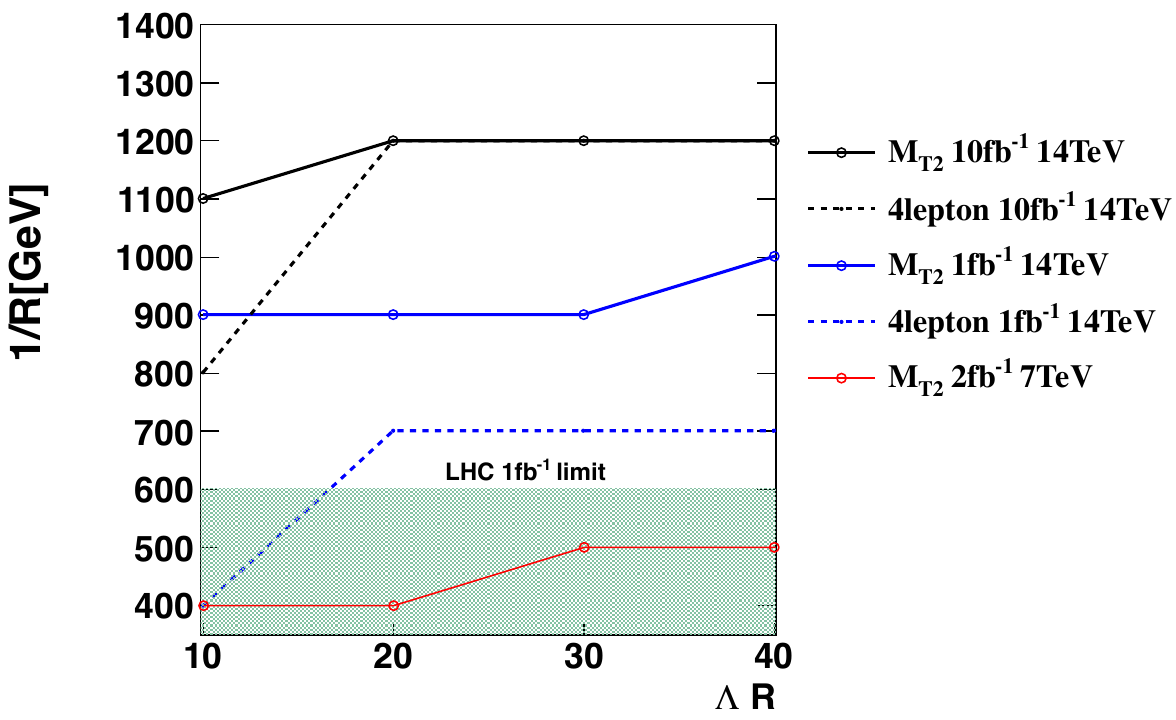,width=4.0in}}
\vspace*{8pt}
\caption{
MUED discovery potential at LHC14 in the 4l+$E_T^{miss}$ analysis.  
From Ref.~\protect\cite{Murayama:2011hj}. 
\protect\label{fig4l}}
\end{figure}

\begin{figure}[!t]
\centerline{\psfig{file=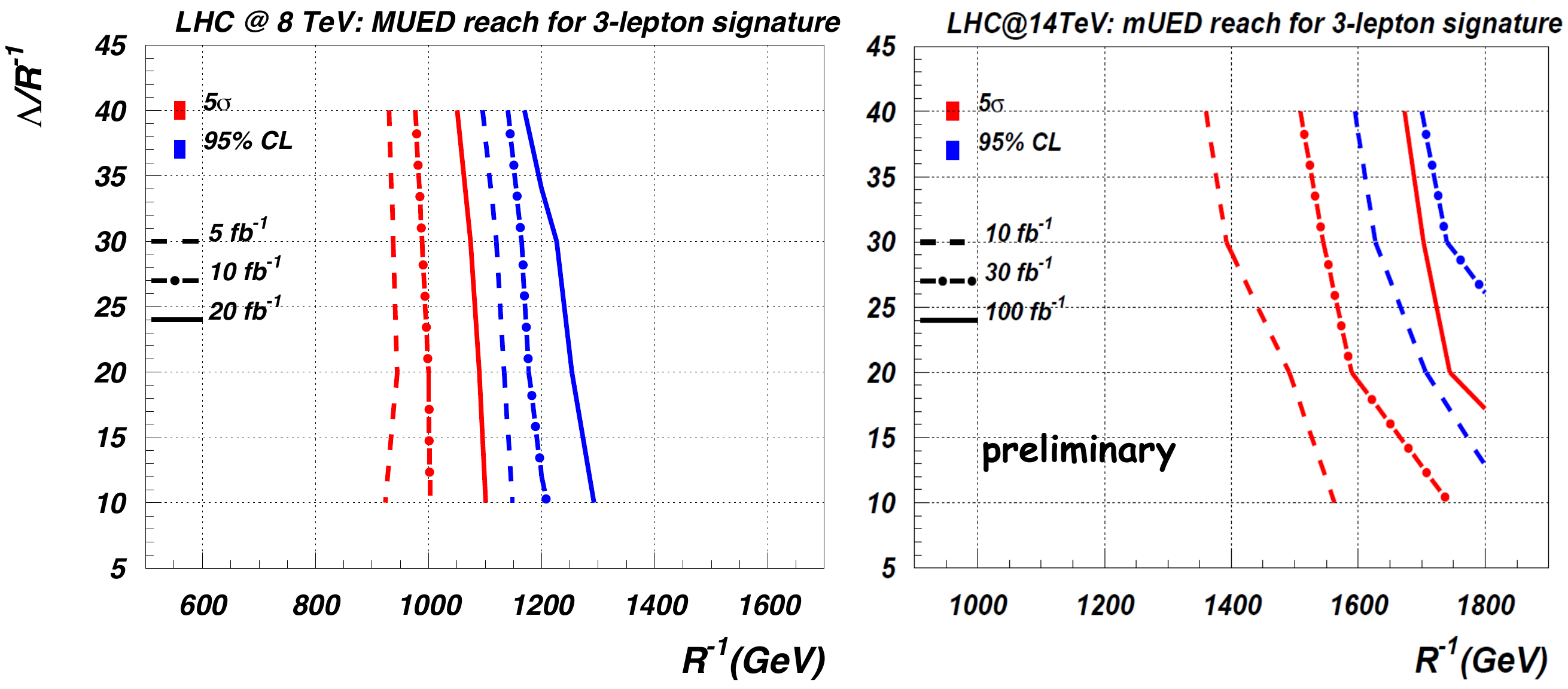,width=6.0in}}
\vspace*{8pt}
\caption{MUED exclusion and discovery potential from the 3-lepton channel at LHC8 (left)\protect\cite{Belyaev:2012ai} and LHC14 (right)~\protect\cite{Belyaev:SUSY2013}.
\protect\label{fig3l}}
\end{figure}

A study of the 4-lepton channel~\cite{Murayama:2011hj}  shows that  10 fb$^{-1}$ of LHC data at 14 TeV enable  to reach  $R^{-1} \sim 1.2$ TeV, see Fig.~\ref{fig4l}. This mass range is still lower than the mass relevant for Dark Matter, however, it was argued that with improvements in the analysis one may hope  to probe the relevant DM region with ${\cal O}$(100) fb$^{-1}$. In the meantime, the 3-lepton channel was proven to be much more effective\cite{Belyaev:2012ai}. As illustrated in Fig.~\ref{fig3l}, the 8 TeV run can already 
set the most stringent bound so far on MUED to $\sim 1.2$ TeV and it is expected that the entire relevant region will be probed with the 14 TeV run~\cite{Belyaev:SUSY2013}.
 There have been no dedicated searches for UED by ATLAS or CMS, and the limits in Ref.~\cite{Murayama:2011hj,Belyaev:2012ai,Belyaev:SUSY2013} are estimates by 
theorists. On the other hand, the  latest ATLAS  searches for new physics in events with three charged leptons \cite{TheATLAScollaboration:2013cia} can be recasted as UED limits. We should note that the conclusions above were derived assuming values of $\Lambda R\sim{\cal O}(10)$. 
Considering very low values of $\Lambda R$~\cite{Kakuda:2013kba,Cornell:2014jza} changes 
the prospects: the KK spectrum is highly degenerate, and although the DM region corresponds to masses beyond 1 TeV,
 the final state leptons become very soft so that
 the LHC sensitivity in the 3 lepton channel is likely to drop.
One has to re-assess the prospects for probing the DM region in this case.

Limits from recent LHC searches for dilepton resonances were derived as well in Ref.~\cite{Edelhauser:2013lia}. The dilepton final state 
can arise from  the production of second-level KK states which can decay into SM particles and thus do not result in any missing transverse momentum.
These analysis lead to the weaker bounds $R^{-1}\gtrsim 715$ GeV on the compactification scale  and $M_{{KK}^{(2)}}\gtrsim 1.4$ TeV on the masses of the second-level  KK particles in the MUED model.
This is consistent with earlier studies ~\cite{Datta:2005zs,Bhattacherjee:2010za}  which showed that the discovery reach at the LHC for level-2 KK modes in MUED, in particular for
$\gamma^{(2)}, Z^{(2)} $ and  $g^{(2)}$,  is not promising in view of the present 
constraints from the 3-lepton channel.

The Higgs searches also lead to constraints on UED models. Due to  additional KK loops, in particular the KK top, Higgs production is enhanced in UED while the Higgs to diphoton branching fraction also receives contributions from the KK W and is suppressed with respect to the SM prediction \cite{Petriello:2002uu}.
However, the corresponding bounds on the KK mass scale are much weaker than those from direct production of KK quarks and gluons, of the order of 600 GeV for MUED\cite{Belanger:2012mc,Kakuda:2013kba}.

\section{Beyond the Minimal UED model}

Mild departures from the minimal UED model lead to significantly different phenomenology. 
A simple extension  is to include boundary interactions localized  on the fixed points of the $S_1/Z_2$ orbifold at the cutoff scale $\Lambda$\cite{Flacke:2008ne}. The next-to-minimal UED (NMUED) model considers in addition the presence of bulk mass terms\cite{Flacke:2013pla}. It has two extra parameters in addition to $R$ and $\Lambda$, a common boundary parameter $r$ and a common bulk mass $\mu$ for the third generation quarks, which allows for a slightly lower compactification scale than in MUED, $R^{-1}\gtrsim 500$ GeV from  EW precision tests and Higgs searches \cite{Flacke:2013nta}. 
It remains to be seen what are the constraints from LHC searches and how they fit with DM predictions.
In fact, in the presence of non-trivial boundary localized terms, the LKP may rather be the first-level $Z$ boson, $Z^1$, or the first-level neutral Higgs boson, $H^1$, see Ref.~\cite{Flacke:2008ne}. The phenomenology of $Z^1$ dark matter was studied in Ref.~\cite{Arrenberg:2008wy,Melbeus:2011gs}. Its preferred mass is typically larger than $B^1$, consequently it is more challenging to detect. $H^1$ is also a viable WIMP candidate but with even worse detection prospects~\cite{Melbeus:2012wi}, essentially due to its suppressed couplings to light fermions.

Another extension is to consider UED models with two extra dimensions~\cite{Appelquist:2002ft,Dobrescu:2004zi,Burdman:2005sr,Ponton:2005kx,Burdman:2006gy}. This construction was motivated as 
 the  number of matter generations can be derived  as a consequence of anomaly cancellation \cite{Dobrescu:2001ae}. Another special property of 6D UED with compactification of the two extra dimensions on a $T_2/Z_4$ orbifold of equal radii (the so-called ``chiral square") is that proton decay is suppressed to acceptable levels even for a baryon number violation scale in the TeV range, as a consequence of  the combination of standard-model gauge invariance and 6-dimensional Lorentz invariance \cite{Appelquist:2001mj,Appelquist:2002ft}.
Theories with two compact universal extra dimensions also contain a KK parity. 4D particles are labelled by two positive integers $(n,m)$. These particles are odd under KK parity when $n+m$ is odd.
In 6D, there are two spin-0 fields transforming in the adjoint representation of the gauge group. One linear combination is eaten to become the longitudinal degree of freedom of the spin-1 KK particle. The other linear combination remains a physical spin-0 particle \cite{Dobrescu:2004zi,Burdman:2005sr,Ponton:2005kx,Burdman:2006gy}.

Neglecting effects from cutoff scale physics localized at the corners of the square compactification, the (1,0) mode of the spinless adjoint of the hypercharge gauge group turns out to be the lightest one, the so-called ``spinless photon", and is stable because of Kaluza-Klein parity. 
Annihilation of this Ôspinless photonÕ was found to proceed predominantly through Higgs boson 
%and is largely independent of other Kaluza-Klein particles. 
exchange  in Ref.~\cite{Dobrescu:2007ec}. According to the estimate of Ref.~\cite{Dobrescu:2007ec},
the measured relic abundance sets an upper limit on the spinless photon mass in the range  $[150-200]$ GeV for a Higgs boson mass around 125 GeV. The phenomenology of this dark matter candidate is strikingly different from the  KK photon of 5-dimensional UED. In particular,  the spinless photon annihilation into fermions is helicity-suppressed and it mainly annihilates into  $W$, $Z$ gauge bosons and Higgs. 
This dark matter candidate resembles more to a neutralino than to a KK photon.
 Indirect detection with gamma rays and antimatter is difficult because of the helicity-suppression of the annihilation cross section into fermions.
It has no spin-dependent scattering cross section and its spin-independent scattering cross section is also helicity suppressed.
This makes indirect detection in neutrino telescopes (due to scattering in the Sun) almost hopeless and direct detection in underground detectors very challenging.

LHC constraints on this  ``chiral square" 6D UED model were investigated in Ref.~\cite{Choudhury:2011jk} which concentrated on a final state comprising multiple jets with a single hard photon and missing energy, arising from the pair production of (1,0) KK gluons and KK quarks. A bound of $\sim 700$ GeV  was derived using 7 TeV LHC data.
This is definitely  in tension with the leading order DM mass estimate of Ref.~ \cite{Dobrescu:2007ec} but it remains to be seen by how much the mass prediction  from the relic abundance calculation would be pushed up  when taking into account higher KK modes, as it happens in MUED.

An alternative compactification on a
 two-dimensional orbifold of the ``Real Projective Plane" (the so-called RP$^2$) was studied in detail as it guarantees an exact KK parity independently of the assumptions about UV scale physics\cite{Cacciapaglia:2009pa}. The KK mass spectrum has smaller mass splittings than in other UED constructions so that coannihilation processes play a crucial role in the DM relic abundance calculation. 
 The DM mass is predicted  to be in the [350-500]  GeV range~\cite{Arbey:2012ke}  in the case of equal radii $R_5=R_6$, which is 
 excluded by direct detection experiments.
 In the asymmetric and hierarchical case, the phenomenology is dominated by one mass scale $R^{-1}=min(1/R_5,1/R_6)$, and the preferred DM mass range is $R^{-1}\sim $[700-1000] GeV. This is safe from
 the LHC bound $R^{-1} \gtrsim $ 600 GeV found in  Ref.~\cite{Cacciapaglia:2013wha} using
ATLAS and CMS searches for events with missing transverse energy,
 but borderline with the direct detection limits, especially for small cutoff scale $\Lambda$.
 
For a discussion of other 6D UED constructions and a more complete list of references, see Ref.~\cite{Kakuda:2013kba} (bounds from DM and from KK mode searches at the LHC  are lacking for these additional models).

%
%\begin{comment}
%\begin{table}[!h]
%\begin{center}
%\begin{tabular}{lcc}
%\hline
%&DM mass range&LHC bound \\
%5D MUED&$R^{-1} \in $ [1300-1500] GeV\cite{Belanger:2010yx} &$R^{-1} \gtrsim $1300 GeV %\cite{Belyaev:2012ai} \\
%{6D UED}, $T^2/Z_4$&$  R^{-1}  \in  [150-200]$ GeV\cite{Dobrescu:2007ec}& $R^{-1} \gtrsim $ 500 GeV %\cite{Choudhury:2011jk} \\
%6D UED, $RP^2$ &$ R^{-1}  \in  [350-500] $ GeV\cite{Arbey:2012ke} & $R^{-1} \gtrsim $ 600 GeV %\cite{Cacciapaglia:2013wha} \\
%\hline
%\end{tabular}
%\end{center}
%\caption{Summary   } 
%\label{tab:summary}
%\end{table}
%\end{comment}

%\begin{figure}[!h]
%\centerline{\psfig{file=Figures/Table1.pdf,width=5.0in}}
%\vspace*{8pt}
%\caption{Comparison of lower bounds on $R^{-1}$ for MUED and various 6D UED models,
%from Ref.~\protect\cite{Kakuda:2013kba}
%\protect\label{figbeyond}}
%\end{figure}

\section{Conclusion}

While offering peculiar LHC phenomenology which will be tested at the next LHC run, 5D UED models bring a still-viable spin-1 WIMP dark matter candidate. At present, apart from Ref.~\cite{ArkaniHamed:2000hv,Appelquist:2002ft,Burdman:2006jj}, there has been no attempt to address EW symmetry breaking in the UED construction
and it remains to be shown how a natural solution to keep the Higgs mass light can be compatible with the underlying assumptions (especially KK number  conservation and KK parity)
 that make the specificities of the UED phenomenology~\cite{Agashe:2007jb}. 
 
 Anyhow, as defined, the 5D MUED model is safe from experimental constraints. The 14 TeV run at the LHC will enable to probe the $R^{-1} \sim 1300-1500$ GeV mass region relevant for dark matter for $\Lambda R\sim {\cal O}(10)$.  
 For $\Lambda R\lesssim 5$, the mass splitting between KK particles is small, leading to very soft final state leptons and therefore a reduced efficiency. 
 A detailed study is required to estimate the LHC prospects for probing the DM region in this case.
 Variants to the MUED model are numerous. Each of them is subject to different constraints. 
Including bulk and boundary mass terms is a natural and testable possibility.
As for  6D UED models, the DM relic density calculation typically predicts a mass range for the spinless 
photon which is in tension with the LHC or other constraints. 

%\vspace{0.25cm}
%{\bf Note}: {Among the extensive bibliography on UED phenomenology, we only referred, for conciseness, in %this mini-review, to the works that lead to the strongest constraints.}

\section*{Acknowledgments}

I thank Genevi\`eve B\'elanger,  Sasha Belyaev, Jesus Moreno  for their feedback, and especially Mitsuru Kakizaki for providing Fig.~\ref{figdirect} and Bogdan Dobrescu for reading the manuscript.

%\section*{References}

\end{document}